\documentclass[twoside]{article}
\usepackage{Proc_NTSE}

\newcommand{\comment}[1]{}
\begin{document}
\thispagestyle{plain}

\begin{center}
{\Large \bf \strut
Accelerating Ab Initio Nuclear Physics Calculations with GPUs
\strut}\\
\vspace{10mm}
{\large \bf

Hugh Potter$^{a}$, Dossay Oryspayev$^{b, c}$, Pieter
Maris$^{a}$, Masha Sosonkina$^{c, d}$ James Vary$^{a}$, Sven Binder$^{e}$,
Angelo Calci$^{e}$, Joachim Langhammer$^{e}$, Robert Roth$^{e}$,
\"{U}mit \c{C}ataly\"{u}rek$^{f}$, Erik
Saule$^{g}$}
\end{center}

\noindent{\small $^a$\it Department of Physics and Astronomy, Iowa State University, Ames, IA 50011, USA} \\
{\small $^b$\it Department of Electrical and Computer Engineering, Iowa State University, Ames, IA 50011, USA} \\
{\small $^c$\it Ames Laboratory, Ames, IA 50011, USA} \\
{\small $^d$\it Department of Modeling, Simulation and Visualization Engineering, Old Dominion University, Norfolk, VA 23529, USA} \\
{\small $^e$\it Institut f\"{u}r Kernphysik - Theoriezentrum, Technische Universit\"{a}t Darmstadt, 64289 Darmstadt, Germany} \\
{\small $^f$\it Departments of Biomedical Informatics, Electrical and Computer Engineering, and Computer Science and Engineering, Ohio State University, Columbus, OH 43210, USA} \\
{\small $^g$\it Department of Computer Science, University of North Carolina at Charlotte, Charlotte, NC 28223, USA}

\markboth{
Hugh Potter et al.}
{
Accelerating Ab Initio Nuclear Physics Calculations with GPUs}

\begin{abstract}
This paper describes some applications of GPU acceleration in ab initio
nuclear structure calculations. Specifically, we discuss GPU acceleration of
the software package MFDn, a parallel nuclear structure eigensolver. We modify
the matrix construction stage to run partly on the GPU. On the Titan
supercomputer at the Oak Ridge Leadership Computing Facility, this produces a speedup of approximately $2.2$x -- $2.7$x for
the matrix construction stage and $1.2$x -- $1.4$x for the entire run.
\\[\baselineskip]
\end{abstract}

\section{Introduction}
The Configuration Interaction approach to computational nuclear physics casts the Schr\"{o}dinger equation for the nuclear many-body bound state problem as a matrix eigenvalue problem \cite{Navratil2000,Vary2009}. The many-body Hamiltonian is approximated by a finite matrix whose eigenvalues and eigenvectors correspond to the bound state energies and wavefunctions.  The wavefunctions can then be used to calculate other observables.  Typically only the lowest eigenvalues and eigenvectors of this matrix are of interest.

The Hamiltonian matrices required to accurately calculate nuclear properties can be very large, with dimension in excess of $10^9$, and $10^{13}$ or more nonzero matrix elements \cite{Vary2009,Maris2012}\comment{ \cite{ref2, ref3} }. Calculations of this magnitude can only be performed on supercomputers, with parallel codes like Many-Fermion Dynamics for nuclei (MFDn) \cite{Sternberg2008,Maris2013,Maris2010,Aktulga2012,Aktulga2013}\comment{ \cite{ref4, ref5, ref6, ref7, ref8} }, a hybrid MPI/OpenMP software package written in Fortran and C.

The Hamiltonian matrices are very sparse, but their sparsity structure is nontrivial; locating and calculating the nonzero matrix elements takes a significant fraction of the overall runtime in MFDn, on the order of $25\%$ -- $45\%$ for some representative cases. The matrix construction stage contains a number of parallelizable steps, however, as each nonzero element can be calculated independently. This problem structure is a promising target for acceleration on SIMD-style coprocessors like GPUs. We present an investigation of GPU acceleration in the matrix construction stage of MFDn.

GPU accelerators pair a large number of cores with a communal block of memory.
They have much less memory per core and cannot easily handle more complex
program logic, but are capable of running many calculations in parallel. The
CUDA framework \cite{NVIDIA2012} from NVIDIA
provides a high-level API for accessing GPU functionality, allowing GPUs to be programmed in languages like C and Fortran.

Code to be executed on the GPU is written in a function called the kernel.
The CPU code can then invoke the kernel, specifying a number of cores on which to run simultaneously. The kernel invocation
specifies thread count in a two-level hierarchy: Threads are grouped together into blocks, and blocks are grouped into a grid.
Each thread has its own small, private, local memory, and each thread in a
block can access the shared memory of that block.
Every thread in the grid has access to the global memory,
which can be in the $1$GB -- $6$GB range.
The user calls CUDA allocation and copy functions to move data to the global
memory of the GPU, invokes the kernel and waits for completion, and then
uses copy functions to retrieve the results of the calculation.
The kernel invocation will often request more threads than the GPU has cores.
In this case, the GPU has a scheduler to stream blocks to cores as they become
available.

\section{Overview of MFDn}
MFDn is a hybrid MPI/OpenMP parallel software package written in Fortran and C
for ab initio nuclear physics calculations. MFDn generates a many-body nuclear
Hamiltonian matrix for the nucleus in question and uses the Lanczos algorithm to extract
the lowest eigenvalues and eigenvectors. The many-body matrix is stored in memory on
core; this strategy limits the sizes of the matrices that can be used, but is
much faster than accessing the matrix from disk. The matrix is symmetric, so
only half of it is generated and stored \cite{Aktulga2012,Aktulga2013}.

MFDn runs in several stages. After the various indexing systems are set up
to specify the many-body basis, the many-body Hamiltonian matrix is constructed. The nonzero elements are then located, calculated, and stored. Elements in the many-body matrix are built up as linear combinations of kinetic energy and nuclear interaction terms. MFDn reads the kinetic energy and 2-body and 3-body potentials from file, and uses them to calculate elements of the many-body Hamiltonian. The many-body matrix is distributed among MPI processes in a way that produces a roughly uniform distribution of nonzero elements.

Once the matrix is generated, MFDn obtains the lowest eigenvalues and eigenvectors with the Lanczos algorithm, an iterative algorithm that relies on successive matrix-vector multiplications and orthogonalizations. The Lanczos algorithm requires many iterations, and is the most computationally-intensive stage. Efficient multicore approaches have been implemented in \cite{Maris2010, Aktulga2012,Aktulga2013}\comment{ \cite{ref7, ref8}}. Performance with respect to non-uniform memory access (NUMA) architecture in supercomputer nodes has been studied in \cite{Srinivasa2012}\comment{ \cite{ref12} }. When the Lanczos algorithm has completed, MFDn uses the eigenvalues and eigenvectors to calculate other observables, which can then be compared to experiment.

Despite being computationally intensive, the Lanczos algorithm stage is not an easy target for GPU acceleration; it is memory-bound, and also cannot be easily broken down into GPU-parallelizable pieces. In the matrix construction stage, however, each many-body matrix element can be calculated independently. Furthermore, in the current implementation of MFDn, each 3-body matrix element that is needed in the many-body matrix must be obtained by performing a change of basis on the input 3-body potential.  This part of the code is very computationally intensive, and we implement GPU acceleration at the level of this basis transformation.

\section{Standalone Basis Transformation on the GPU}
\subsection{Basis Transformation Algorithm}
One method for storing the $3$-body input interaction matrix is to use the
coupled-$JT$ basis, which adds, or ``couples'' the angular momenta of the three Single Particle States (SPSs) together into one total angular momentum for the $3$-body state. Isospin, a quantum number that has to do with whether a nucleon is a proton or a neutron, is similarly coupled. This basis exploits the rotational symmetry of the interaction to reduce the amount of information that must be stored.  However, for constructing the many-body matrix elements, we need 3-body interaction matrix elements in an $m$-scheme basis; that is, we need to ``decouple'' these coupled-$JT$ matrix elements every time we need a 3-body matrix element in the construction of the many-body matrix.  Storing the 3-body interaction matrix in $m$-scheme would be more efficient for the calculations, but requires much more memory: in one representative case, a $3$-body interaction is $33$GB in the $m$-scheme basis, but only $1$GB in the coupled-$JT$ basis. Substantial memory savings can thus be achieved by storing the input matrix in-core in the coupled-$JT$ basis, and calculating $m$-scheme elements individually as they are required by MFDn \cite{Maris2013,Roth2011,Roth2013}\comment{ \cite{ref5, ref13, ref14} }.

\begin{figure}
\centerline{\includegraphics[width=0.8\textwidth]{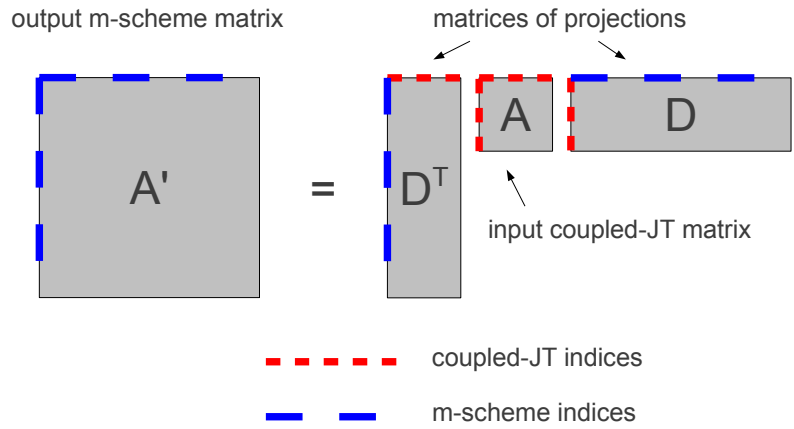}}
\caption{Matrix $A$ is transformed from the coupled-$JT$ basis to the $m$-scheme basis through multiplication with $D$ and $D^T$.}
\label{matrix_multiply}
\end{figure}

When MFDn requires a $3$-body input interaction matrix element, then, it must convert that element from the coupled-$JT$ basis. From linear algebra, basis transformations of matrices are of the form $A'=D^TAD$, where $D$ is a matrix of projections from one basis to the other. Figure \ref{matrix_multiply} shows a high-level illustration of this transformation. As in any basis transformation, an element in the new basis is a linear combination of elements from the old basis, weighted by the projections in $D$.

\begin{figure}
\centerline{\includegraphics[width=0.8\textwidth]{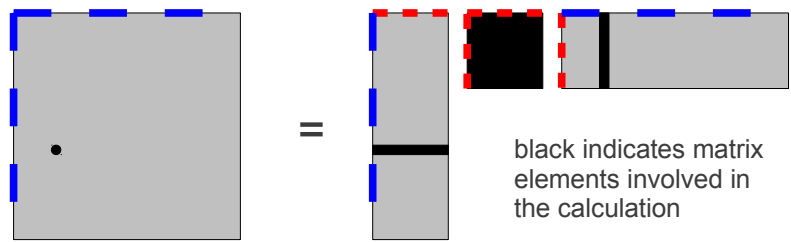}}
\caption{MFDn requests one 3-body element at a time.  The single indicated element in $A'$ can be calculated from the indicated elements in $A$, $D$, and $D^T$.  In practice, not all the indicated elements are used, as orthogonality relations dictate that many of them are zeroes.}
\label{matrix_multiply_black}
\end{figure}

In the coupled-$JT$ to $m$-scheme transformation, $D$ is developed from a series of angular momentum and isospin coupling coefficients. The matrices $A$, $D$, and $D^T$ are never actually constructed in their entirety. MFDn requests $3$-body elements from $A$ one-at-a-time, and elements of $D$ are developed as needed for each request; Figure \ref{matrix_multiply_black} illustrates the calculation of a single element from $A'$.

In principle an element of $A'$ is a linear combination of all elements from $A$. Many elements from $A$ do not contribute, however, because of orthogonality relations that manifest as zeroes in $A$ and $D$. This sparsity structure is highly predictable and can be exploited. $A$ and $A'$ can be divided into blocks in a way that allows any element in a block in $A$ to be constructed entirely from elements in the corresponding block in $A'$. Furthermore, all the nonzero elements within that block can be iterated over with a set of nested loops over coupled angular momentum and isospin values.

Only these potentially-nonzero elements are stored, and they are arranged in the order that the nested loops will reference them. The basis conversion routine, then, consists of locating the start of the correct block, going through the nested angular momentum and isospin loops, and adding the elements of $A$ one after the other, each weighted by coupling coefficients calculated from the corresponding angular momentum and isospin values. In practice, because the relevant isospin space is so small, the isospin coupling coefficients are precalculated with several conditionals, and the isospin loops are unrolled into a single weighted summation. The core of the calculation, then, is a set of three nested loops over coupled angular momentum values.

\subsection{GPU Implementation}
The bounds of the inner angular momentum decoupling loop depend on the positions in the outer loops, and the bounds of the outer loops depend on which $3$-body element is being calculated. The loop structure is thus irregular. This irregularity makes it difficult to map GPU threads to parts of the problem when dedicating more than one thread to the calculation of a single $3$-body element. We parallelized with one $3$-body element calculation per GPU thread, invoking the kernel across many $3$-body element calculations at once.

We used CUDA to interface with the GPU. We put the nested loop structure into the kernel without modifying it significantly, and added a wrapper function to transfer a chunk of $3$-body element requests to the GPU, invoke the kernel, and transfer the results back. MFDn and the decoupling code flatten the SPS quantum numbers into a single linear SPS index, so a $3$-body element request takes the form of a set of six SPS indices. We also flattened several arrays that were multidimensional in the CPU-only code so that they could be transferred to and referenced on the GPU more quickly.

Before integrating this GPU acceleration strategy into MFDn, we applied it to a standalone version of the basis transformation code to test performance and act as an intermediate step; we observed a $4$x to $10$x speedup compared to a multithreaded CPU implementation running on eight cores \cite{Oryspayev2013}\comment{ \cite{ref15} }.

\section{Integration into MFDn}
\subsection{A Closer Look at Matrix Construction}
In the matrix construction stage of MFDn, the many-body Hamiltonian is divided into chunks of elements with similar quantum numbers, and these chunks are apportioned across MPI processes. Each process then splits into OpenMP threads to count, locate, and calculate the nonzero many-body elements. Elements are stored per MPI process in a single array, with a list of column pointers to split them into columns and a secondary array to denote their locations in their columns (compressed row format, or CSR).

The nonzero elements are located in one of two ways. If a block is denser, all elements in it are iterated through and tested for being nonzero. If the block is sparser, all combinations of quantum numbers that could yield nonzero elements are iterated through. MFDn runs through nonzero elements twice during the many-body matrix generation: once to count the nonzero matrix elements so the appropriate arrays can be allocated, and once to calculate them.

\subsection{Integration with Standalone GPU Code}
During the construction of the many-body matrix,  MFDn uses a recursive loop to calculate the nonzero many-body matrix elements, frequently requesting 3-body matrix elements in $m$-scheme. In
the CPU-only version of MFDn, the decoupling code calculates $3$-body elements
one-by-one, as they are requested. The GPU version of the decoupling code
requires a large block of simultaneous requests to be efficient, so the
sequential requests in the CPU code are not ideal. To bridge the gap between
MFDn and the GPU decoupling code, we use buffers to store lists of $3$-body
element requests so large ``chunks'' of requests can be sent to the GPU at once.

Each OpenMP thread has its own buffer allocated to store requests. On
receiving a request, the CPU part of the decoupling subroutine stores the
request in the buffer and returns $0$. In the CPU-only version, the returned
value is added directly to the many-body element under calculation; we must
thus also store which many-body element the request pertains to so that it can
be added to the correct many-body element when the calculation finishes on the
GPU. Furthermore, the $3$-body element is added with a specific phase, which
must also be stored with the request.  Larger buffers are more efficient
because the overhead associated with the single CUDA memory copy and kernel
invocation is split over more elements.  Tests with the standalone code
indicate diminishing returns after around 20,000 elements \cite{Oryspayev2013}
on the supercomputer Dirac at the National Energy Research Scientific
Computing Center. Hence, we use here buffers of approximately this size, although further testing may be required to optimize buffer size for the integrated code and for different hardware.

Each OpenMP thread starts in the so-called ``accumulating mode'' while passing
element requests to its buffer until it is full.  Then, the thread sends the
buffer to the GPU and switches into the so-called ``non-accumulating mode''. In this mode, the decoupling code runs as in the CPU-only version of MFDn, calculating $3$-body elements on the CPU at request and returning them; this allows the CPU to continue work while the GPU, which may be shared among many OpenMP threads, is busy. The thread checks periodically for a completed chunk from the GPU. When it receives the chunk, it iterates through the returned three-body elements in the chunk, multiplies them by the stored phases, and adds them to the array of many-body elements at the stored locations. It then switches back into accumulating mode, and the cycle begins again. At the end of the many-body matrix construction phase, all the $3$-body contributions have been added in, either from the GPU calculations or directly from the CPU decoupling code.

\section{Technical Concerns}
\subsection{MPI and OpenMP Structure of GPU Access}
We allow the GPU to decide which requests to calculate first. Each OpenMP thread is given its own stream to access the GPU, so at any given time the GPU will have a number of requests open. With this scheme, a single GPU may end up being accessed by multiple MPI processes. This is not possible on all systems, and incurs a performance penalty when it is possible.

One alternative is to restrict the number of MPI processes to the number of
nodes used (one per node).  The resulting MPI/OpenMP divisions, however, may not be ideal for the
NUMA structure of the node, which may incur a performance
penalty due to data locality and thread contention
issues~\cite{Srinivasa2012}.  Another alternative is to GPU-accelerate only some MPI
processes, leaving the others to run as in the CPU-only version. However,
using this alternative would derange the almost perfect load-balancing, which
is a prominent design feature of the CPU-only version of MFDn, and thus,
diminish the benefit of the acceleration.  Given these considerations, we opt
to accept the performance penalty from multiple MPI processes per GPU on
systems where that is possible, and restrict the number of MPI processes to
the number of nodes on other systems.

\subsection{Indexing the Input Interaction}
The decoupling code requires the use of a six-dimensional index array, which
relates quantum numbers to locations in the input coupled-$JT$ interaction
matrix. This array is highly jagged: the length along any particular dimension
is not constant, but rather depends on the position along the other
dimensions. In C, jagged arrays are represented as trees of pointers, wherein
the ``root'' pointer points to an array of pointers, each of which point to
further arrays of pointers, and thus until the ``leaves'', which hold the actual data of the array.

Generating such a structure requires a prodigious number of allocations, as each array at each level must be allocated. On the CPU this is not an issue, but allocations on the GPU must first be requested from the CPU, adding a significant transfer time penalty. Our first attempt did not take this inefficiency into account, and over $90\%$ of the matrix construction time was spent generating the index array on the GPU.

To address this problem, we generated the entire index array in a contiguous block on the CPU, producing a structure wherein the pointers were correct relative to each other. We then applied a constant offset to all the pointers in the index array so that their absolute coordinates would be correct for a contiguous block allocated on the GPU. The entire pointer structure could then be copied into that block with one copy, allowing the index structure to be created on the GPU with a single GPU allocate and GPU copy. With this improvement the index array creation time becomes negligible in the overall matrix construction.

\section{Initial Experiments}
We present results from the DOE supercomputer Titan at the Oak Ridge National
Laboratory.  Titan is a Cray XK7 supercomputer with 18,688 physical compute
nodes, each of which has one 16-core 2.2 GHz AMD Opteron 6274 processor and 32
GB of RAM.  Each node is divided into two NUMA domains, and nodes are served
in groups of two by Gemini high-speed interconnect routers.  Additionally,
each node has one NVIDIA K20 Tesla GPU accelerator with 6 GB of memory.

We use the number of non-zeroes in the many-body matrix as a measure of problem size, and test at a variety of problem sizes.  The number of nonzeroes is determined by the choice of nucleus and truncation, so it is difficult to provide a smooth spectrum.  Different problem sizes can require vastly different numbers of cores to store the many-body matrix, so we do not test all problems sizes on the same configuration; for each problem size, we allow the many-body matrix to take up half of the total memory, and choose the smallest configuration that satisfies that requirement.  We implement and test GPU acceleration in MFDn version 14, and compare against CPU performance using an unmodified build of version 14.

\begin{figure}
\centerline{\includegraphics[width=0.8\textwidth]{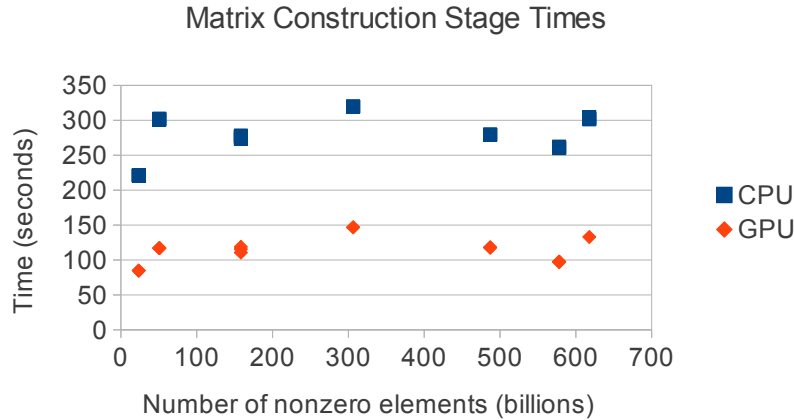}}
\caption{Matrix construction times with CPU-only and GPU-accelerated code.}
\label{construction_time}
\end{figure}

\begin{figure}
\centerline{\includegraphics[width=0.8\textwidth]{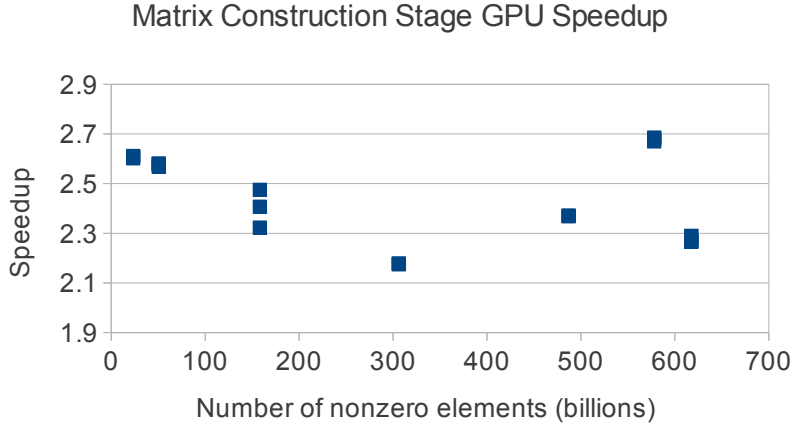}}
\caption{Matrix construction stage speedup with GPU acceleration.}
\label{speedup}
\end{figure}


Our primary results are summarized in Figures \ref{construction_time} and
\ref{speedup}: We see a speedup of 2.2x~--~2.7x in the matrix construction
stage.  There is no immediately apparent pattern in the dependence of speedup
on problem size.  The choices of nuclei and truncation parameters required to
generate the spectrum of problem sizes are somewhat haphazard, so it is
possible that any problem size dependence has become entangled with dependence
on those parameters.  Despite the individually varying speedup, the range of
speedups appears to stay roughly the same; thus the GPU acceleration appears
to scale well for the problem sizes investigated.


The speedup over the entire run is a more ambiguous quantity.  The time taken
in the MFDn diagonalization stage depends on how many eigenvalues are required
and the accuracy to which they are required to converge.  The speedup over the
entire run, which depends on the relative times of the matrix construction and
diagonalization stages, therefore depends on these parameters also.  For the
representative parameter choices used in the matrix construction speedup
calculations, the overall speedup is in the 1.2x -- 1.4x range.

\section{Conclusion}
We have modified the matrix construction stage of MFDn to run partly on the GPU. The current MFDn implementation stores the matrix elements of the $3$-body input interaction in the compressed coupled-$JT$ basis in-core. The conversion of these elements back to $m$-scheme for use with MFDn is highly parallelizable; we implement this basis transformation on the GPU.

Initial timing results with the GPU-accelerated MFDn code are promising.  We
have achieved a consistent speedup in the two- to three-fold range for the
matrix construction stage, and  our speedup scales smoothly to larger problem
sizes, at least for those investigated in this paper.  It may be possible,
though more difficult, to leverage GPU acceleration in the diagonalization
stage, or at a higher level in the matrix construction stage; such
improvements are left for future consideration.

\section{Acknowledgements}
This work was supported in part by Iowa State University under the contract
DE-AC02-07CH11358 with the U.S. Department of Energy (DOE), by the U.S. DOE
under the grants DESC0008485 (SciDAC/NUCLEI), and DE-FG02-87ER40371
(Division of Nuclear Physics), by the Director, Office of Science, Division
of Mathematical, Information, and Computational Sciences of the U.S.
Department
of Energy under contract number DE-AC02-05CH11231, and in part by the National
Science Foundation grant NSF/OCI-0941434, 0904782, 1047772.
This research was also supported (in part) by the DFG through SFB 634 and by
HIC for FAIR.

This research used resources of the Oak Ridge Leadership Computing Facility at the Oak Ridge National Laboratory, which is supported by the Office of Science of the U.S. Department of Energy under Contract No. DE-AC05-00OR22725, and resources of the National Energy Research Scientific Computing Center, which is supported by the Office of Science of the U.S. Department of Energy under Contract No. DE-AC02-05CH11231.


\end{document}